\documentclass[10pt]{article}
\usepackage{amssymb}
\usepackage{amsmath}
\usepackage{amsfonts}
\usepackage{amsthm}
\usepackage{amsxtra}
\usepackage{graphicx}
\usepackage{txfonts}
\newlength{\mytopmargin}
\newlength{\myleftmargin}
\newtheorem{thm}{Theorem}[section]
\newtheorem{prop}[thm]{Proposition}
\newtheorem{lem}[thm]{Lemma}

\theoremstyle{remark}
\newtheorem{rem}{Remark}[section]
\newcommand{\ol}[1]{\overline{#1}}
\newcommand{\ul}[1]{\underline{#1}}
\newcommand\Arg[1]{{\rm Arg}~{#1}}
\newcommand{\DS}[1]{\displaystyle{#1}}
\title{\bf Hypergeometric Solutions to the $q$-Painlev\'e Equation of
Type $(A_1+A_1')^{(1)}$}
\author{Taro Hamamoto${}^{1}$, Kenji Kajiwara${}^{1}$ and Nicholas S. Witte${}^2$\\[3mm]
{\normalsize ${}^1$ Graduate School of Mathematics, Kyushu University,
6-10-1 Hakozaki, Fukuoka 812-8581, Japan}\\
{\normalsize ${}^2$ Department of Mathematics and Statistics,
University of Melbourne, Victoria 3010, Australia}\\
}
\date{July 27, 2006}
\begin{document}
\maketitle
\begin{abstract}
A class of classical solutions to the $q$-Painlev\'e equation of type
$(A_1+A_1')^{(1)}$ (a $q$-difference analog of the Painlev\'e II equation) is
constructed in a determinantal form with basic hypergeometric function
elements. The continuous limit of this $q$-Painlev\'e equation to the
Painlev\'e II equation and its hypergeometric solutions are
discussed. The continuous limit of these hypergeometric solutions to the
Airy function is obtained through a uniform asymptotic expansion of their
integral representation.
\end{abstract}
\section{Introduction}
In this article we consider the following $q$-difference equation
\begin{equation}
 \left(\ol{F}F-1\right)\left(F\ul{F}-1\right)=\frac{at^2F}{F+t},\label{eq:qp2}
\end{equation}
where $t$ is the independent variable
\begin{equation}
\ol{t}=qt,\quad \ul{t}=t/q,\quad F=F(t),\quad \ol{F}=F(qt),\quad \ul{F}=F(t/q),
\end{equation}
with $|q|<1$ and $a$ is a parameter. Equation (\ref{eq:qp2}) was identified as 
one of the $q$-difference Painlev\'e equations by Ramani and
Grammaticos\cite{RG:coales} with a continuous limit to the Painlev\'e II
equation (P$_{\rm II}$). In this sense eq.(\ref{eq:qp2}) is sometimes
regarded as a $q$-analog of P$_{\rm II}$. Sakai formulated the discrete 
Painlev\'e equations as Cremona transformations on a certain family of rational
surfaces and developed their classification
theory\cite{Sakai:elliptic}. According to this theory eq.(\ref{eq:qp2})
is a discrete dynamical system on the rational surface characterized by
the Dynkin diagram of type $A_6^{(1)}$, which posseses the symmetry of the
affine Weyl group of type $(A_1+A_1')^{(1)}$. Equation (\ref{eq:qp2})
may be denoted as dP($A_6^{(1)}$) by the notation adopted in \cite{MSY}.
Although eq.(\ref{eq:qp2}) is the simplest nontrivial
$q$-difference Painlev\'e equation that admits a B\"acklund
transformation only a few results are known - its continuous
limit\cite{RG:coales} and its simplest hypergeometric
solution\cite{RGTT,KMNOY:hyper1,KMNOY:hyper2}. The first purpose of this
article is to construct ``higher-order'' hypergeometric solutions to
eq.(\ref{eq:qp2}) explicitly in determinantal form.

The second purpose of this article is to consider the continuous limit
in some detail. The limiting procedure works well on the formal level of the
defining $q$-difference equation however na\"ive application of the
procedure does not work on the level of their solutions. The application
of the continuous limit to the series representation of the basic
hypergeometric functions that appear in the solutions does not yield the
Airy functions which are the hypergeometric solutions of P$_{\rm II}$.
To obtain the valid limit we follow the procedure used by
Prellberg\cite{Prellberg} - we construct an appropriate integral
representation of the function and derive an asymptotic expansion by
applying a generalization of the saddle point method.

Our paper is organized as follows. In section 2 we construct
hypergeometric solutions to eq.(\ref{eq:qp2}). The simplest solution is
obtained in section 2.1, and determinant formula of ``higher-order'' solutions
is presented in section 2.2, whose proof is given in section 2.3. In
section 3 we consider the continuous limit as $q\to 1^{-}$. The limit on the 
formal level is discussed in section 3.1. We discuss the limit on the level of
hypergeometric functions in section 3.2. Section 4 is devoted to
concluding remarks.

\section{Hypergeometric Solutions and Their Determinant Formula}
\subsection{The Simplest Solution}
The simplest hypergeometric solution to eq.(\ref{eq:qp2})
is obtained by looking for the special case where it reduces to the
Riccati equation. Then by linearizing the Riccati equation we obtain
a second order linear $q$-differential equation, which admits
basic hypergeometric functions as solutions. 

Let us first recall the definition of the basic hypergeometric
series\cite{GR:2004}
\begin{equation}
 {}_r\varphi_s\left(\begin{array}{c}a_1,\ldots,a_r
		 \\b_1,\ldots,b_s\end{array};q,z\right)
=\sum_{n=0}^\infty\frac{(a_1,\ldots,a_r;q)_n}{(b_1,\ldots,b_s;q)_n(q;q)_n}
\left[(-1)^nq^{\binom{n}{2}}\right]^{1+r-s}z^n,
\end{equation}
where
\begin{equation}
 (a_1,\ldots,a_r;q)_n=\prod_{i=1}^n(a_i;q)_n,\quad (a;q)_n=(1-a)(1-qa)\cdots(1-q^{n-1}a).
\end{equation}
Then the simplest hypergeometric solution to eq.(\ref{eq:qp2}) is given
as follows (see also \cite{KMNOY:hyper1,KMNOY:hyper2,RGTT}):
\begin{lem}\label{lem:simplest}
Eq.(\ref{eq:qp2}) admits the following particular solution for $a=q$:
\begin{equation}
F=\frac{\ol{\psi}}{\psi},\label{eq:dep}
\end{equation}
where $\psi(t)$ satisfies the linear $q$-difference equation
\begin{equation}
\ol{\psi}+t\psi=\ul{\psi}.\label{eq:qAiry}
\end{equation}
The general solution of eq.(\ref{eq:qAiry}) is given by
\begin{equation}
 \psi(t)=A~{}_1\varphi_1\left(\begin{array}{c}0 \\-q \end{array};q,-qt\right)+B~
{\rm e}^{\pi i\frac{\log t}{\log q}}~{}_1\varphi_1\left(\begin{array}{c}0 \\-q \end{array};q,qt\right),
\label{eq:gen_sol_of_qAiry}
\end{equation}
where $A$ and $B$ are arbitrary $q$-periodic functions.
\end{lem}
\noindent{\it Proof.} It is easy to see that if $F$ is the solution of the Riccati equation 
\begin{equation}
\ol{F}=\frac{1}{F}-qt,\label{eq:Riccati}
\end{equation}
then $F$ satisfies eq.(\ref{eq:qp2}) with $a=q$.  Equation
(\ref{eq:Riccati}) can be linearized via eq.(\ref{eq:dep}) to
eq.(\ref{eq:qAiry}) by putting $F=\phi/\psi$ and equating the numerators and
denominators of both sides.

We next substitute $\psi=t^\rho\sum\limits_{n=0}^\infty a_nt^n$ into
eq.(\ref{eq:qAiry}). Then we have $q^{2\rho}=1$, which implies
$\rho=\frac{m\pi i}{\log q}$ $(m\in\mathbb{Z})$. Furthermore we deduce a recursion
relation for $a_n$ from eq.(\ref{eq:qAiry})
\begin{displaymath}
 a_n=(-1)^m\frac{q^n}{1-q^{2n}}a_{n-1}.
\end{displaymath}
Accordingly we obtain two fundamental 
solutions to eq.(\ref{eq:qAiry}) as
\begin{displaymath}
 \psi_1(t)=\sum_{n=0}^\infty
  \frac{(-1)^nq^{n(n-1)/2}}{(q;q)_n(q;-q)_n}~(-qt)^{n},\quad
 \psi_2(t)={\rm e}^{\pi i\frac{\log t}{\log q}}\sum_{n=0}^\infty
\frac{(-1)^nq^{n(n-1)/2}}{(q;q)_n(q;-q)_n}~(qt)^{n},
\end{displaymath}
for $m=0,1$ respectively.
This completes the proof of Lemma \ref{lem:simplest}. $\square$

\subsection{B\"acklund Transformation and Determinant Formula}
Sakai constructed the following transformations for the homogeneous
variables $x$, $y$, $z$ and the parameters $a_0$, $a_1$, $b$ on the $A_6^{(1)}$
type ({\sl Mul.7}) surface\footnote{Actions of these transformations are modified from
the original formulae in \cite{Sakai:elliptic} so that they are
subtraction-free. Some typographical errors have been also fixed.}:
\begin{equation}
\begin{array}{l}
\sigma:\quad(a_1,a_0,b\ ;\ x:y:z)\mapsto (a_0,a_1,a_1b\ ;\ z(z+x): bx(z+x):yz)\\
\sigma_{(15)(24)(60)}:\quad(a_1,a_0,b\ ;\ x:y:z)\\
\hskip50pt \mapsto (1/a_0,  1/a_1,   b\ ;\ z(x+z)(x+y+z): y((z+bx)(x+z)+yz): bx(x+z)^2)\\
w_1:\quad (a_1,a_0,b\ ;\ x:y:z)\mapsto (1/a_1,  a_1^2a_0,  a_1b\ ;\ x(y+a_1z) : y(y+z) : a_1z(y+z))\\
w_0=  \sigma_{(15)(24)(60)}w_1\sigma_{(15)(24)(60)}.
\end{array}
\end{equation}
The action of $\sigma^2$
\begin{equation}
\sigma^2:\quad (a_1,a_0,b\ ;\ x:y:z)
\longmapsto 
(a_1,a_0,a_0a_1b\ ;\ yz^2(x+y+z): a_1bz^2(x+y+z)(x+z):bxyz(x+z)).\label{eq:sigma2}
\end{equation}
gives rise to eq.(\ref{eq:qp2}) by putting
\begin{equation}
a=\left(\frac{a_0}{a_1}\right)^{1/2},\quad 
t = \left(\frac{b^2}{a_0^3a_1}\right)^{1/4},
\quad q=(a_0a_1)^{1/2},
\qquad F=\left(\frac{a_0}{a_1b^2}\right)^{1/4}~\frac{y}{x}. \label{eq:qp2var}
\end{equation}
Note that these transformations satisfy the fundamental relations,
\begin{displaymath}
w_0^2=w_1^2=1,\quad w_0\sigma^2 = \sigma^2 w_0,\quad w_1\sigma^2=\sigma^2 w_1.
\end{displaymath}
Then $w_0$ and $w_1$ can be regarded as B\"acklund transformations
of eq.(\ref{eq:qp2}). In particular, the action of $T=\sigma w_1$ is
given by
\begin{equation}
 T(a)=q^2 a,\quad T(t)=t,\quad T(F)=\frac{qat\ol{F}+\ol{F}F-1}
{(\ol{F}F-1)(t\ol{F}+\ol{F}F-1)}.
\end{equation}
Therefore applying $T$ to the ``seed'' solution in Lemma
\ref{lem:simplest}, we obtain ``higher-order'' hypergeometric
solutions to eq.(\ref{eq:qp2}) expressible in terms of
a rational function of $\psi$ for $a=q^{2N+1}$ ($N\in\mathbb{Z}$).  It is
observed that the numerators and denominators of such solutions are
factorized and those factors admit the following Casorati determinant
formula.
\begin{thm}\label{thm:det}
For each $N\in\mathbb{Z}$, we define $\tau_N(t)$ by
\begin{equation}
\tau_N(t)=
\left\{
\begin{array}{cl}\bigskip
    \begin{vmatrix}
    \psi(t) & \psi(q^{2}t) & \cdots & \psi(q^{2N-2}t) \\
    \psi(q^{-1}t) & \psi(qt)  & \cdots & \psi(q^{2N-3}t) \\
    \vdots & \vdots & & \vdots\\
    \psi(q^{-N+1}t) & \psi(q^{-N+3}t) & \cdots & \psi(q^{N-1}t)
   \end{vmatrix} & (N > 0),\\
\bigskip
   1 & (N=0),\\
   \begin{vmatrix}
    \psi(q^{-1}t) & \psi(q^{-3}t) & \cdots & \psi(q^{-2M+1}t) \\
    \psi(t) & \psi(q^{-2}t)  & \cdots & \psi(q^{-2M+2}t) \\
    \vdots & \vdots & & \vdots\\
    \psi(q^{M-2}t) & \psi(q^{M-4}t) & \cdots & \psi(q^{-M}t)
   \end{vmatrix} & (M=-N,\ N<0).
\end{array}
\right.\label{tau:qp2}
\end{equation}
Then 
\begin{equation}
F(t)=
\left\{
\begin{array}{ll}
\medskip
{\displaystyle \frac{1}{q^{N}}~\frac{\tau_{N}(t)\tau_{N+1}(qt)}{\tau_{N}(qt)\tau_{N+1}(t)}}  
&(N\geq 0), \\
{\displaystyle-\frac{1}{q^{N+1}}~\frac{\tau_{N}(t)\tau_{N+1}(qt)}{\tau_{N}(qt)\tau_{N+1}(t)}}  
& (N<0),
\end{array}
\right.\label{eq:dep_var_transf}
\end{equation}
satisfies eq.(\ref{eq:qp2}) with $a = q^{2N+1}$.
\end{thm}
Theorem \ref{thm:det} is the direct consequence of the following proposition.
\begin{prop}\label{prop:bl}
The $\tau_N$ satisfy the following bilinear $q$-difference equations of
Hirota type:
\begin{enumerate}
 \item $N \geq 0  $
\begin{align}
& q^{2N}\tau_{N+1}(t/q)\tau_{N}(q^{2}t) - q^{N}t~\tau_{N+1}(t)\tau_{N}(qt)
- \tau_{N+1}(qt)\tau_{N}(t)=0,\label{eq:bl1}\\
& q^{2N}\tau_{N+1}(t/q)\tau_{N}(qt) - q^{2N}t~\tau_{N+1}(t)\tau_{N}(t)
- \tau_{N+1}(qt)\tau_{N}(t/q)=0. \label{eq:bl2}
\end{align}
 \item $N < 0$
\begin{align}
&q^{2N+2}\tau_{N+1}(t/q)\tau_{N}(q^{2}t) +
q^{N+1}t~\tau_{N+1}(t)\tau_{N}(qt)-\tau_{N+1}(qt)\tau_{N}(t)=0,
   \label{eq:bl3}\\
&q^{2N+2}\tau_{N+1}(t/q)\tau_{N}(qt)+
q^{2N+1}t~\tau_{N+1}(t)\tau_{N}(t)-\tau_{N+1}(qt)\tau_{N}(t/q)=0.\label{eq:bl4}
\end{align}
\end{enumerate}
\end{prop}
In fact one can derive Theorem \ref{thm:det} from Proposition
\ref{prop:bl} as follows: 
for $N\geq 0$ the bilinear equations eqs.(\ref{eq:bl1}) and (\ref{eq:bl2}) can be rewritten as
\begin{equation}
q^{2N}-\frac{ \mu_{N+1}(t/q) }{\mu_{N}(qt)}
\left(q^{N}t +  \frac{\mu_{N+1}(t)}{\mu_{N}(t)}\right) =0, \quad
q^{2N}-q^{2N}t~\frac{\mu_{N+1}{(t/q)}}{\mu_{N}{(t)}} -
\frac{\nu_{N}{(qt)}}{\nu_{N}{(t/q)}} =0,\label{eq:qp2aux}
\end{equation}
respectively by introducing the variables
\begin{equation}
\nu_{N}(t)=\frac{\tau_{N+1}(t)}{\tau_{N}(t)},\quad
\mu_{N}(t)=\frac{\tau_{N}(qt)}{\tau_{N}(t)}.
\label{eq:nu_and_mu}
\end{equation}
Putting
\begin{equation}
F=\frac{1}{q^N}\frac{\nu_N(qt)}{\nu_N(t)}
=\frac{1}{q^N}\frac{\tau_{N+1}(qt)\tau_N(t)}{\tau_{N+1}(t)\tau_N(qt)},
\end{equation}
and eliminating $\mu_{N}$ and $\mu_{N+1}$ from eq.(\ref{eq:qp2aux})
through the use of the identity
\begin{equation}
 \frac{\nu_{N}(qt)}{\nu_{N}{(t)}}=\frac{\mu_{N+1}(t)}{\mu_{N}(t)},
\end{equation}
we obtain eq.(\ref{eq:qp2}) with $a=q^{2N+1}$. The case of
$N<0$ can be verified in the same way. 

\subsection{Proof of Proposition \ref{prop:bl}}
Our basic idea for proving Proposition \ref{prop:bl} is to use
a determinantal technique. Bilinear $q$-difference equations are derived
from the Pl\"ucker relations which are quadratic identities among
determinants whose columns are shifted. Therefore, we first construct
such ``difference formulas'' that relate ``shifted determinants'' and
$\tau_N$ by using the $q$-difference equation of $\psi$. We then derive
bilinear difference equations with the aid of difference formulas from
proper Pl\"ucker relations. We refer to
\cite{KK:qp3,KNY:qp4,KOS:dP3,KOSGR:dP2,NSKGR:alt-dP2} for 
applications of this method to hypergeometric solutions of other discrete
Painlev\'e equations.

Let us consider the case of $N>0$. We first introduce a notation for
the determinants
\begin{equation}
\tau_N(t) = \begin{vmatrix}
    \psi(t) & \psi(q^{2}t) & \cdots & \psi(q^{2N-2}t) \\
    \psi(q^{-1}t) & \psi(qt)  & \cdots & \psi(q^{2N-3}t) \\
    \vdots & \vdots & & \vdots\\
    \psi(q^{-N+1}t) & \psi(q^{-N+3}t) & \cdots & \psi(q^{N-1}t)
   \end{vmatrix}  =  \left|\ \Psi_{0},\Psi_{2},\ldots,\Psi_{2N-2}\  
\right|,\label{eq:det_notation}
\end{equation}
where $\Psi_{k}$ denotes a column vector
\begin{equation}
\Psi_{k}=\left(\begin{array}{c}\psi(q^kt)\\\psi(q^{k-1}t)\\\vdots\\\psi(q^{k-
N+1}t)\\ \end{array}\right).
\end{equation}
Here the height of the column vector is $N$ however we employ the same symbol for
determinants with differing heights.

\begin{lem}\label{lem:dif}
The following formulas hold:
\begin{align}
&\left|\ \Psi_{0},\Psi_{2},\ldots,\Psi_{2N-2}\ \right|=\tau_N(t),\\
&|\ \widehat{\Psi}_{0}, \Psi_{1} , \Psi_{3} , \ldots , \Psi_{2N-3}\
|=(-1)^{N-1}q^{-(N-1)(N-2)/2}t^{-N+1}~\tau_N(t),\label{eq:dif1}\\
&|\ \Psi_{1} , \widehat{\Psi}_{2} , \Psi_{3} , \ldots , \Psi_{2N-3}\ |
=(-1)^{N-2}q^{-(N-1)(N-2)/2}t^{-N+1}~\tau_N(t),\label{eq:dif2}
\end{align}
where $\widehat{\Psi}_{k}$ denotes the column vector
\begin{equation}
  \widehat{\Psi}_{k}=\begin{pmatrix} \psi(q^{k}t) \\ q\psi(q^{k-1}t) \\ 
\vdots \\
q^{N-1}\psi(q^{k-N+1}t)\end{pmatrix}.
\end{equation}
\end{lem}
\textit{Proof.} 
Using the linear equation eq.(\ref{eq:qAiry}) for $\psi$ on the $N$-th column of
the determinant eq.(\ref{eq:det_notation}) we find
\begin{eqnarray*}
\tau_N(t) &=&
\begin{vmatrix}
    \psi(t) & \psi(q^{2}t) & \cdots & \psi(q^{2N-4}t)
&\psi(q^{2N-4}t)-q^{2N-3}t\psi(q^{2N-3}t)\\
    \psi(q^{-1}t) & \psi(qt)  & \cdots &\psi(q^{2N-5}t) &
\psi(q^{2N-5}t)-q^{2N-4}t\psi(q^{2N-4}t)\\
    \vdots & \vdots & & \vdots\\
    \psi(q^{-N+1}t) & \psi(q^{-N+3}t) & \cdots &\psi(q^{N-3}t) &
\psi(q^{N-3}t)-q^{N-2}t\psi(q^{N-2}t)
   \end{vmatrix},  \\
&=&\begin{vmatrix}
    \psi(t) & \psi(q^{2}t) & \cdots & -q^{2N-3}t\psi(q^{2N-3}t)\\
    \psi(q^{-1}t) & \psi(qt)  & \cdots & -q^{2N-4}t\psi(q^{2N-4}t)\\
    \vdots & \vdots & & \vdots\\
    \psi(q^{-N+1}t) & \psi(q^{-N+3}t) & \cdots & -q^{N-2}t\psi(q^{N-2}t)
   \end{vmatrix}.
\end{eqnarray*}
Applying the same procedure from the $(N-1)$-th column to the second column
we have
\begin{eqnarray*}
   \tau_N(t)&=&
\begin{vmatrix} 
\psi(t)& -qt\psi(qt)& \dots & -q^{2N-3}t\psi(q^{2N-3}t) \\
\psi(q^{-1}t)& -t\psi({t})& \dots & -q^{2N-4}t\psi(q^{2N-4}t) \\
\vdots & \vdots & \quad &\vdots \\
\psi(q^{-N+1}) & -q^{-N+2}t\psi(q^{-N+2}t)& \dots &-q^{N-2}t\psi(q^{N-2}t)
\end{vmatrix}\\
&=&(-t)^{N-1}q^{(N-1)^2}\begin{vmatrix}
    \psi(t) & \psi(qt) & \cdots & \psi(q^{2N-3}t) \\
    \psi(q^{-1}t) & q^{-1}\psi(t)  & \cdots & q^{-1}\psi(q^{2N-4}t) \\
    \vdots & \vdots & & \vdots\\
\psi(q^{-N+1}t) & q^{-N+1}\psi(q^{-N+2}t) & \cdots & q^{-N+1}\psi(q^{N-2}t)
   \end{vmatrix}\\
&=&(-t)^{N-1}q^{(N-1)(N-2)/2}\begin{vmatrix}
    \psi(t) & \psi(qt) & \cdots & \psi(q^{2N-3}t) \\
    q\psi(q^{-1}t) & \psi(t)  & \cdots & \psi(q^{2N-4}t) \\
    \vdots & \vdots & & \vdots\\
q^{N-1}\psi(q^{-N+1}t) & \psi(q^{-N+2}t) & \cdots &\psi(q^{N-2}t)
   \end{vmatrix}\\
&=&(-t)^{N-1}q^{(N-1)(N-2)/2}~|\ \widehat{\Psi}_{0} , \Psi_{1} , \Psi_{3} , \dots , \Psi_{2N-3} \ |,
\end{eqnarray*}
which is nothing but eq.(\ref{eq:dif1}). At the stage where the above procedure
has been employed up to the third column we have
\begin{displaymath}
 \tau_{N}{(t)} =
\begin{vmatrix} 
\psi(t) & \psi(q^{2}t) & -q^{3}t\psi(q^{3}t) &\dots & -q^{2N-3}t\psi(q^{2N-3}t) \\
\psi(q^{-1}t) & \psi(qt) & -q^{2}t\psi(q^{2}t) & \dots & -q^{2N-4}t\psi(q^{2N-4}t) \\
\vdots & \vdots & \vdots & \quad &\vdots \\
\psi(q^{-N+1}t) & \psi(q^{-N+3}t) & -q^{-N+4}t\psi(q^{-N+4}t)& \dots & -q^{N-2}t\psi(q^{N-2}t)
\end{vmatrix}.
\end{displaymath}
Using eq.(\ref{eq:qAiry}) on the first column, we obtain
\begin{eqnarray*}
\tau_{N}{(t)} &=&
\begin{vmatrix}
\psi(q^{2}t)+qt \psi(qt) & \psi(q^{2}t) & -q^{3}t\psi(q^{3}t) &\dots  & -q^{2N-2}\psi(q^{2N-3}t) \\
\psi(qt)+t \psi(t) & \psi(qt) & -q^{2}t\psi(q^{2}t) & \dots & -q^{2N-3}\psi(q^{2N-4}t) \\
\vdots & \vdots & \vdots & \quad &\vdots \\
\psi(q^{-N+3}t) +q^{-N+2}t \psi(q^{-N+2}t) & \psi(q^{-N+3}t) & -q^{-N+4}t\psi(q^{-N+4}t)& \dots & -q^{N-1}\psi(q^{N-2}t)
\end{vmatrix} \\
&=&(-1)^{N-2}q^{(N-1)(N-2)/2}t^{N-1}
\begin{vmatrix}
\psi(qt) &\psi(q^{2}t) &\psi(q^{3}t) &\cdots & \psi(q^{2N-3}t) \\
\psi(t) & q \psi(qt) & \psi(q^{2}t) & \dots & \psi(q^{2N-4}t) \\
\vdots & \vdots & \vdots & \quad &\vdots \\
\psi(q^{-N+2}t) & q^{N-3}\psi(q^{-N+3}t) & \psi(q^{-N+4}t)& \dots & \psi(q^{N-2}t)
\end{vmatrix}\\
&=&(-1)^{N-2}q^{(N-1)(N-2)/2}t^{N-1}|\ \Psi_{1} ,\widehat{\Psi}_{2} ,\Psi_{3} ,\dots ,\Psi_{2N-3}\ | ,
\end{eqnarray*}
which is eq. (\ref{eq:dif2}). $\square$

Now consider the Pl\"ucker relation,
\begin{eqnarray*}
0 &=&\left|\ \Psi_{-1} , \widehat{\Psi}_{0} , \Psi_{1} , \dots , \Psi_{2N-5}\ \right| 
\times \left|\ \Psi_{1} , \Psi_{3} , \dots , \Psi_{2N-3}, \phi\ \right| \\
  &-&\left |\ \widehat{\Psi}_{0} , \Psi_{1} , \Psi_{3} , \dots , \Psi_{2N-5}, \phi\ \right|
 \times \left|\ \Psi_{-1} , \Psi_{1} , \Psi_{3} ,\dots, \Psi_{2N-3} \ \right| \\
 &+& \left|\ \Psi_{-1} , \Psi_{1} , \Psi_{3} , \dots , \Psi_{2N-5} , \phi\ \right| 
\times \left|\ \widehat{\Psi}_{0} , \Psi_{1} , \Psi_{3} ,\dots, \Psi_{2N-3} \ \right|, 
\end{eqnarray*}
for an arbitrary column vector $\phi$. In particular by choosing $\phi$ as
\begin{displaymath}
 \phi=\left(\begin{array}{c} 0\\\vdots\\0\\1\end{array}\right)\quad\mbox{or}\quad
\left(\begin{array}{c} 1\\0\\\vdots\\1\end{array}\right),
\end{displaymath}
and applying Lemma \ref{lem:dif} we obtain eq.(\ref{eq:bl1}) from the
former and eq.(\ref{eq:bl2}) from the latter, respectively. The case of
$N<0$ can be proved in a similar manner. This completes the proof of
Proposition \ref{prop:bl}. $\square$
\section{Continuous Limit to P$_{\rm II}$}
\subsection{Continuous Limit on the Formal Level}
The continuous limit of ``$q$-P$_{{\rm II}}$'' to P$_{{\rm II}}$ involves
the ``quantum'' to classical limit $q\to 1$ but in contrast to the trivial limits
usually employed in basic hypergeometric series, 
i.e. making the substitution $ z \mapsto (1-q)z $ and then setting $ q\to 1 $, we
have a completely different limiting process which is far from trivial and has
not been studied much. Let us first recall the formal limits of
dP($A_6^{(1)}$) eq.(\ref{eq:qp2}).

\begin{prop}\label{prop:limit1}\cite{RG:coales}
With the replacements
\begin{equation}
F=i{\rm e}^{-\delta w},\quad a={\rm e}^{-\frac{\eta}{2}\delta^3},\quad
q={\rm e}^{-\frac{\delta^3}{2}},\quad t=-2i{\rm e}^{-\frac{s}{2}\delta^2}=-2iq^{\frac{s}{\delta}},
\label{eq:cont_params}
\end{equation}
eq.(\ref{eq:qp2}) has a limit to P$_{\rm II}$
\begin{equation}
 \frac{d^2w}{ds^2}=2w^3+2sw+\eta,\label{eq:p2}
\end{equation}
as $ \delta \to 0 $.
\end{prop}
Proposition \ref{prop:limit1} can be easily verified by noticing that
\begin{equation}
 w(q^{\pm 1}t)= w(s\pm\delta)
=w \pm \delta \frac{d}{ds}w + \frac{\delta^2}{2!}\frac{d^2w}{ds^2}+O(\delta^3).\label{eq:wt}
\end{equation}

It is well-known that P$_{\rm II}$ eq.(\ref{eq:p2}) admits the
hypergeometric solutions for $\eta=2N+1$
($N\in\mathbb{Z}$)\cite{Okamoto:p2}:
\begin{equation}
 w=-\frac{d}{ds}\log\frac{\kappa_{N+1}}{\kappa_N},\label{eq:p2_dep_var_transf}
\end{equation}
\begin{equation}
\kappa_N=
\left\{
\begin{array}{cl}
\bigskip
\begin{vmatrix}
v & \frac{\DS{dv}}{\DS{ds}} &\cdots &\left(\frac{\DS{d}}{\DS{ds}}\right)^{N-1}v\\
\frac{\DS{dv}}{\DS{ds}} & \frac{\DS{d^2v}}{\DS{ds^2}} &\cdots &\left(\frac{\DS{d}}{\DS{ds}}\right)^{N}v\\
\vdots & \vdots &\ddots &\vdots\\
\left(\frac{\DS{d}}{\DS{ds}}\right)^{N-1}v &\left(\frac{\DS{d}}{\DS{ds}}\right)^{N}v  &\cdots &\left(\frac{\DS{d}}{\DS{ds}}\right)^{2N-2}v 
\end{vmatrix}& (N>0),\\
\bigskip
1 & (N=0),\\
\begin{vmatrix}
v & \frac{{\DS{dv}}}{{\DS{ds}}} &\cdots &\left(\frac{\DS{d}}{\DS{ds}}\right)^{M-1}v\\
\frac{\DS{dv}}{\DS{ds}} & \frac{\DS{d^2v}}{\DS{ds^2}} &\cdots &\left(\frac{\DS{d}}{\DS{ds}}\right)^{M}v\\
\vdots & \vdots &\ddots &\vdots\\
\left(\frac{\DS{d}}{\DS{ds}}\right)^{M-1}v &\left(\frac{\DS{d}}{\DS{ds}}\right)^{M}v  &\cdots &\left(\frac{\DS{d}}{\DS{ds}}\right)^{2M-2}v 
\end{vmatrix}& (N=-M<0),\\
\end{array}
\right.\label{eq:p2tau}
\end{equation}
where $v$ satisfies the Airy equation,
\begin{equation}
 \frac{d^2v}{ds^2}=-sv.\label{eq:Airy}
\end{equation}
The general solution to eq.(\ref{eq:Airy}) is given by
\begin{equation}
 v(s)=C{\rm Ai}({\rm e}^{\frac{\pi i}{3}}s)+D{\rm Ai}({\rm  e}^{-\frac{\pi i}{3}}s),
\label{eq:gen_sol_of_Airy}
\end{equation}
where $C$, $D$ are arbitrary constants and ${\rm Ai}(s)$ is the Airy
function defined by
\begin{equation}
{\rm Ai}(s)=\frac{1}{2\pi}\int_{-\infty}^\infty {\rm e}^{\frac{i}{3}u^3+isu}~du.
\end{equation}
\begin{prop}
Under the substitutions eq.(\ref{eq:cont_params}) and with
\begin{equation}
 \psi(t)={\rm e}^{\frac{\pi i}{2}\frac{\log t}{\log q}}v(t),
\label{vDefn}
\end{equation}
the hypergeometric solutions to eq.(\ref{eq:qp2}) for
$a=q^{2N+1}$ ($N\in\mathbb{Z}$), eqs.(\ref{eq:qAiry}), (\ref{tau:qp2})
and (\ref{eq:dep_var_transf}) yield, in the limit $\delta\to 0$, the hypergeometric 
solutions to P$_{\rm II}$ eq.(\ref{eq:p2}) for $\eta=2N+1$, eqs. (\ref{eq:Airy}),
(\ref{eq:p2tau}) and (\ref{eq:p2_dep_var_transf}).
\end{prop}
\noindent\textit{Proof.} 
Noticing that 
\begin{displaymath}
\psi(q^{\pm 1}t)=\pm i {\rm e}^{\frac{\pi i}{2}\frac{\log t}{\log q}}v(s\pm\delta),
\end{displaymath}
the linear equation eq.(\ref{eq:qAiry}) can then be rewritten as
\begin{equation}
 v(s+\delta)+v(s-\delta)=2{\rm e}^{-\frac{s}{2}\delta^2}v(s),
\end{equation}
which yields eq.(\ref{eq:Airy}) in the limit $\delta\to 0$.
We next consider the limit of $\tau_N$ and the dependent variable
transformation. The determinant $\tau_N$ ($N>0$) can be rewritten as
\begin{eqnarray*}
 \tau_N&=&{\rm e}^{\frac{N\pi i}{2}\frac{\log t}{\log q}}
\begin{vmatrix}
    v(t) & i^2v(q^{2}t) & \cdots & i^{2N-2}v(q^{2N-2}t) \\
    i^{-1}v(q^{-1}t) & iv(qt)  & \cdots & i^{2N-3}v(q^{2N-3}t) \\
    \vdots & \vdots & & \vdots\\
    i^{-N+1}v(q^{-N+1}t) & i^{-N+3}v(q^{-N+3}t) & \cdots & i^{N-1}v(q^{N-1}t)
   \end{vmatrix}\\
&=&{\rm e}^{\frac{N\pi i}{2}\frac{\log t}{\log q}} i^{N(N-1)/2}
\begin{vmatrix}
    v(s) & v(s+2\delta) & \cdots & v(s+(2N-2)\delta) \\
    v(s-\delta) & v(s+\delta)  & \cdots & v(s+(2N-3)\delta) \\
    \vdots & \vdots & & \vdots\\
    v(s-(N-1)\delta) & v(s-(N-3)\delta) & \cdots & v(s+(N-1)\delta)
   \end{vmatrix}\\
&=&{\rm e}^{\frac{N\pi i}{2}\frac{\log t}{\log q}}~i^{N(N-1)/2}~\sigma_N(s),
\end{eqnarray*}
so that
\begin{equation}
 F=\frac{i}{q^N}\frac{\sigma_N(s)\sigma_{N+1}(s+\delta)}{\sigma_N(s+\delta)\sigma_{N+1}(s)}.
\end{equation}
We also note that
\begin{displaymath}
\sigma_N(s)=(-2\delta^2)^{N(N-1)/2}~\left[\kappa_N+O(\delta)\right].
\end{displaymath}
Therefore we deduce
\begin{equation}
 w=-\frac{1}{\delta}\log \frac{F}{i}=-\frac{1}{\delta}\log\left[
1+\delta\left(\frac{d}{ds}\log\frac{\kappa_{N+1}}{\kappa_N}\right)+O(\delta^2)
\right]
=-\frac{d}{ds}\log\frac{\kappa_{N+1}}{\kappa_N}+O(\delta).
\end{equation}
The limit in the case of $N<0$ can be verified in a similar manner. $\square$

Now let us consider the limit of the solution of the linear equation
eq.(\ref{eq:qAiry}).  The two linearly independent power-series solutions
of the Airy equation eq.(\ref{eq:Airy}) are given by
\begin{equation}
\begin{array}{l}
\medskip
{\displaystyle {}_0F_1\left(\begin{array}{c} -\\ \frac{2}{3} \end{array} ;
 -\frac{s^3}{3^2}\right)
=1-\frac{1}{3!}s^3 + \frac{1\cdot 4}{6!}s^6 -\frac{1\cdot 4\cdot 7}{9!}s^9+\cdots,}\\
{\displaystyle s~{}_0F_1\left(\begin{array}{c}- \\ \frac{4}{3} \end{array} ;
	     -\frac{s^3}{3^2}\right)
=s-\frac{2}{4!}s^4 + \frac{2\cdot 5}{7!}s^7 -\frac{2\cdot 5\cdot 8}{10!}s^{10}+\cdots.}
\end{array}\label{eq:hyper}
\end{equation}
However as is apparent from the series expansion of hypergeometric functions in
eq.(\ref{eq:gen_sol_of_qAiry}) the application of the scaling changes of
variables in eq.(\ref{eq:cont_params}) does not yield any meaningful limit as
$\delta \to 0$ on a term by term basis.  What is required is another
representation of these functions and a uniform, possibly asymptotic, expansion
with respect to the other parameters as $ q \to 1 $. This question has been
addressed in \cite{Prellberg} and for the most part answered there. We discuss
the continous limit of the hypergeometric functions 
${}_1\varphi_1\left(\begin{array}{c}0 \\-q \end{array};q,\mp qt\right)$ 
in the next section.
\subsection{Continuous Limit of the Hypergeometric Functions}
There are three key ingredients in \cite{Prellberg} which are necessary
to derive the final formula that we require. The first is a suitable
integral representation for the 
${}_1\varphi_1\left(\begin{array}{c}0 \\y \end{array};q,x\right)$ 
function and this is the $q$-analog of the Mellin-Barnes inversion integral.
\begin{prop}\cite{GR:2004,Prellberg} \label{Prop:MBint}
The representation
\begin{equation}
   {}_1\varphi_1\left({{0}\atop{y}} ; q,x\right)
   = \frac{(q;q)_{\infty}}{(y;q)_{\infty}}
     \int_{\rho-i\infty}^{\rho+i\infty}
\frac{dz}{2\pi i}~z^{-\log x/\log q}~
\frac{(y/z;q)_{\infty}}{(z;q)_{\infty}},
\label{eq:MB_qInt}
\end{equation}
is valid for $ x,y \in \mathbb{C} $, with $ |{\rm arg}(x)|<\pi $, $ y \neq q^{-n} $ 
($ n \in \mathbb{Z}_{\geq 0} $), $0<\rho<1$ and $ 0<q<1 $.
\end{prop}
\begin{rem}
Proposition \ref{Prop:MBint} follows by evaluating the integral on the
contour described in Fig.1 with the residues at $z=q^{-n}$
\begin{equation}
  {\mathop{\rm Res}\limits_{z=q^{-n}}}(z;q)^{-1}_{\infty}
  = (-1)^{n+1}\frac{q^{{{n}\choose{2}}}}{(q;q)_n(q;q)_{\infty}}, \quad n \in \mathbb{Z}_{\geq 0} ,
\label{eq:q-residue}
\end{equation}
and by deforming the path $C$ appropriately according to Cauchy's Theorem. 
\begin{figure}[htbp]
\begin{center}
\includegraphics{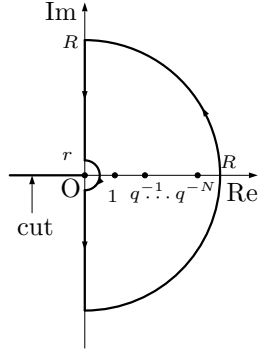}
\end{center}
\caption{
%Left: integration path for eq.(\ref{eq:MB_qInt}). Right:
Contour for Proposition \ref{Prop:MBint}.}
\end{figure} 
\end{rem}
The second ingredient is an asymptotic formula for the $q$-shifted
factorial $(t;q)_{\infty} $ as $ q \to 1$ which is uniform with respect
to $t$. Such expansions have only recently
been studied and in particular by Meinardus \cite{Meinardus:1954}, McIntosh
\cite{McIntosh:1995,McIntosh:1998,McIntosh:1999} and the above cited
work \cite{Prellberg}. Amongst all the essentially equivalent forms we
choose the following statement:
\begin{prop}\cite{Meinardus:1954,Prellberg,McIntosh:1999}\label{Prop:q-factorial}
As $ q \to 1^{-} $ the $q$-shifted factorial $(t;q)_{\infty} $ has an asymptotic
expansion
\begin{equation}
 \log (t;q)_{\infty} \sim
 \frac{1}{\log q}{\rm Li}_2(t) + \frac{1}{2}\log(1-t)
 + \sum^{\infty}_{n=1}\frac{B_{2n}}{(2n)!}\left(t\frac{d}{dt}\right)^{2n-2}\frac{t}{1-t}
   (\log q)^{2n-1} ,
\label{eq:qFac_exp}
\end{equation}
for $ 0<q<1 $ and uniform for $ t$ in any compact domain of $\mathbb{C}$ such 
that $ |{\rm arg}(1-t)|<\pi $. Here $ {\rm Li}_2(t) $ is the dilogarithm
function defined by
\begin{equation}
  {\rm Li}_2(t)=-\int_0^t \frac{\log(1-u)}{u}~du=\sum_{n=1}^\infty \frac{t^n}{n^2},
\end{equation}
and $B_{2n}$ the even Bernoulli numbers.
In the case of $t=q$ we have \cite{Hardy}
\begin{equation}
\log(q;q)_\infty=\frac{\pi^2}{6\log q}+\frac{1}{2}\log\frac{2\pi}{-\log
  q}+O(\log q).
\end{equation}
\end{prop} 
We next apply Proposition \ref{Prop:q-factorial} to the integral
representation eq.(\ref{eq:MB_qInt}). Noticing that $q$-shifted
factorials in the integral representation are rewritten by putting $q={\rm e}^{-\epsilon}$ as
\begin{equation*}
 \frac{(y/z;q)_\infty}{(z;q)_\infty}z^{-\log x/\log q}
  ={\rm e}^{\frac{1}{\epsilon}\left[-{\rm Li}_2(y/z)+{\rm Li}_2(z)+\log x\log z\right]}
\times {\rm
e}^{\frac{1}{2}\left[\log\left(1-\frac{y}{z}\right)-\log(1-z)\right]}\times [1+O(\epsilon)],
\end{equation*}
and
\begin{equation*}
 \frac{(q;q)_\infty}{(y;q)_\infty}
  ={\rm e}^{\frac{1}{\epsilon}\left[{\rm Li}_2(y)-\frac{\pi^2}{6}\right]
  +\frac{1}{2}\log\frac{2\pi}{\epsilon}-\frac{1}{2}\log(1-y)}\times [1+O(\epsilon)],
\end{equation*}
we obtain:
\begin{prop} \label{Prop:Laplace_qInt}
Let $x,y\in\mathbb{C}$ and $q={\rm e}^{-\epsilon}$ for $\epsilon>0$. Then
\begin{equation}\label{eq:Laplace_qInt}
   {}_1\varphi_1\left({{0}\atop{y}} ; q,x\right)=\frac{1}{2\pi i}\int_{\rho-i\infty}^{\rho+i\infty}
{\rm e}^{\frac{1}{\epsilon}\left[\log x\log z-{\rm Li}_2(\frac{y}{z})+{\rm Li}_2(z)\right]}
{\rm e}^{\frac{1}{2}\left[\log(1-\frac{y}{z})-\log(1-z)\right]}~dz
\times \left[1+O(\epsilon)\right],
\end{equation}
where $|\arg(x)|<\pi$, ${\rm Re}~y<\rho$.
\end{prop} 
\begin{rem}
We take $ \log z $ on its principal sheet cut along $ (-\infty,0] $ and $ {\rm Li}_2(z) $
on its principal sheet cut along $ (1,+\infty) $.
If $ x,y \in (0,1) $ then for $ z \in (y,1) $ the argument of 
\begin{equation}
   \log x\log z-{\rm Li}_2(\frac{y}{z})+{\rm Li}_2(z)
\label{gz}
\end{equation}
is zero. When $ z \in \mathbb{C} $ subject to  
\begin{displaymath}
 |\arg z|<\pi, \quad |\arg(1-z)|<\pi , \quad |\arg \left(1-\frac{y}{z}\right)|<\pi ,
\end{displaymath}
that is we exclude the rays $ (1,\infty) $, $ (-\infty,0) $ and $ (0,y) $,
it follows that the argument of eq.(\ref{gz}) lies in the interval $ (-\pi,\pi ) $. The 
contour path given in Proposition \ref{Prop:Laplace_qInt} is then just a simple path 
$ z=\rho+it $ ($ {\rm Re}~y<\rho<1 $, $ t \in (-\infty,\infty) $) satisfying these 
criteria and the requirement that the endpoints of the contour ensure the 
existence of the integral. If the contour is
deformed then eq.(\ref{eq:Laplace_qInt}) is valid if the contour does
not cut across the ray $(0,y)$.
\end{rem}

The third ingredient is the application of saddle point method to the
Laplace type integral in eq.(\ref{eq:Laplace_qInt}). In this problem two
saddle points arise and can coalesce depending on the values of the
parameters. Therefore we have to construct an asymptotic approximation
that incorporates the contributions from both saddle points uniformly 
with repsect to their separation. A method for such an asymptotic expansion of 
this type of integral has been set out by
Chester, Friedman and Ursell\cite{CFU:1957,Ursell:1965,Wong:book}.

%A cubic polynomial arises because there are only two saddle points and the vanishing of
%the exponent at the coalescence point has a linear dependence, and this is the reason
%why the Airy and related functions arise. 

To illustrate the method, let us consider the integral
\begin{displaymath}
 I = \frac{1}{2\pi i}\int_C {\rm e}^{\frac{1}{\epsilon}g(z;d)}f(z)~dz,\quad 
  \epsilon\to 0,
\end{displaymath}
where $f(z)$ is analytic with respect to $z$, and $g(z;d)$ is analytic
with respect to $z$ and the parameter $d$. We assume that there are two
saddle points $z_1$ and $z_2$ which are determined from $g'(z;d)=0$ and
that they coalesce when $d=0$. The key of this method is to introduce
the change of variable $z\to u$ via the cubic parameterization
\begin{equation}\label{eq:g_and_u}
g(z;d)=\frac{1}{3}u^3-\alpha u+\beta,
\end{equation}
where $\alpha$ and $\beta$ are determined as follows. Firstly differentiating
eq.(\ref{eq:g_and_u}) we have
\begin{equation}\label{eq:g_diff}
g'(z;d)\frac{dz}{du}=u^2-\alpha.
\end{equation}
In order for eq.(\ref{eq:g_and_u}) to define a single-valued analytic
transformation neither $\frac{\DS{dz}}{\DS{du}}$ nor $\frac{\DS{du}}{\DS{dz}}$ can vanish
in relevant regions. Therefore at the saddle points we have the
correspondence
\begin{equation}\label{eq:saddle_points_correspondence}
z= z_{1}\ \leftrightarrow u=\alpha^{\frac{1}{2}},\quad
z= z_{2}\ \leftrightarrow u=-\alpha^{\frac{1}{2}},
\end{equation}
which determines $\alpha$ and $\beta$ as
\begin{equation}\label{eq:alpha_beta}
\frac{4}{3}\alpha^{\frac{2}{3}} = g(z_2;d)-g(z_1;d), \quad
2\beta = g(z_2;d)+g(z_1;d).
\end{equation}
The transformation $u=u(z;d)$ defined by eqs.(\ref{eq:g_and_u}) and
(\ref{eq:alpha_beta}) has three branches. However, it can be shown that
there is exactly one branch which has the following
properties\cite{CFU:1957,Wong:book}; (i) $u=u(z;\alpha)$ is expanded
into a power series in $z$ with coefficient continuous in $d$ near $d=0$,
(ii) $z_1$ and $z_2$ correspond to $\alpha^{\frac{1}{2}}$ and
$-\alpha^{\frac{1}{2}}$ respectively, and (iii) near $d=0$ the correspondence
$z\leftrightarrow u$ is 1:1.

We next expand $f(z)$ in the form
\begin{equation}\label{eq:f_expansion}
  f(z)\frac{dz}{du}=\sum_{m=0}^\infty (p_m+q_mu)(u^2-\alpha)^m,
\end{equation}
and define the following integrals
\begin{equation}
\begin{array}{l}\medskip
{\displaystyle F_m=\frac{1}{2\pi i}\int_{C'} 
{\rm e}^{\frac{1}{\epsilon}\left(\frac{1}{3}u^3-\alpha u\right)}(u^2-\alpha)^m~du,} \\
{\displaystyle G_m=\frac{1}{2\pi i}\int_{C'}
{\rm e}^{\frac{1}{\epsilon}\left(\frac{1}{3}u^3-\alpha
 u\right)}(u^2-\alpha)^mu~du,}
\end{array}
\quad
m=0,1,\ldots.
\end{equation}
Here $C'$ is the image of $C$ by the transformation given by
eqs.(\ref{eq:g_and_u}) and (\ref{eq:alpha_beta}). By using recursion
relations for $F_m$ and $G_m$ obtained by partial integration, the
expansion of $I$ can be written in the form
\begin{equation}\label{eq:expansion_Airy}
I ={\rm e}^{\frac{\beta}{\epsilon}}\left[
\epsilon^{\frac{1}{3}}V(\alpha\epsilon^{\frac{2}{3}})\sum_{m=0}^\infty
 a_m\epsilon^m
+\epsilon^{\frac{2}{3}}V'(\alpha\epsilon^{\frac{2}{3}})\sum_{m=0}^\infty
 b_m\epsilon^m
\right],
\end{equation}
where
\begin{equation}
a_0=p_0,\quad b_0=-q_0,
\end{equation}
and $V(\lambda)$ is the Airy integral
\begin{equation}
 V(\lambda)=\frac{1}{2\pi i}\int_{C'} {\rm e}^{\frac{1}{3}u^3-\lambda u}du.
\end{equation}
The coefficients $p_0$ and $q_0$ are determined by putting $z=z_1$,
$u=\alpha^{\frac{1}{2}}$ and $z=z_2$, $u=-\alpha^{\frac{1}{2}}$ in
eq.(\ref{eq:f_expansion}) as
\begin{equation}\label{eq:p0_and_q_0}
\begin{array}{l}
 {\displaystyle 
p_0=\frac{1}{2}\left[f(z_1)\left(\frac{dz}{du}\right)_{z=z_1}+f(z_2)\left(\frac{dz}{du}\right)_{z=z_2}\right],}\\
{\displaystyle q_0=\frac{1}{2\alpha^{\frac{1}{2}}}\left[
f(z_1)\left(\frac{dz}{du}\right)_{z=z_1}-f(z_2)\left(\frac{dz}{du}\right)_{z=z_2}\right].}
\end{array}
\end{equation}
Prellberg \cite{Prellberg} has applied the above expansion to the
integral representation eq.(\ref{eq:Laplace_qInt}) to obtain its leading
behaviour as $\epsilon\to 0$ for $0<x,y<1$ as follows:
\begin{prop}\cite{Prellberg}\label{prop:1-phi-1}
Let $ 0< x,y < 1 $ and $ q=e^{-\epsilon} $ with $ \epsilon>0 $. Then as $\epsilon\to 0$
we have
\begin{eqnarray}
  {}_1\varphi_1\left({0\atop y};q,x\right)
  &=& {\rm e}^{\frac{1}{\epsilon}[{\rm Li}_2(y)-\frac{1}{6}\pi^2+\frac{1}{2}\log x\log y]}
      \sqrt{\frac{2\pi}{\epsilon(1-y)}}\nonumber\\
&&\times \left[ p_0 \epsilon^{\frac{1}{3}}{\rm Ai}(\alpha\epsilon^{-\frac{2}{3}})
           - q_0 \epsilon^{\frac{2}{3}}{\rm Ai}'(\alpha\epsilon^{-\frac{2}{3}}) \right]
           \left[1+{\rm O}(\epsilon) \right],\label{eq:1-phi-1_asymptotic}
\end{eqnarray}
where the auxiliary variables are
\begin{equation}
  \frac{4}{3}\alpha^{3/2} = \log x\log\left(\frac{z_m-\sqrt{d}}{z_m+\sqrt{d}}\right)
    +2{\rm Li}_2(z_m-\sqrt{d})-2{\rm Li}_2(z_m+\sqrt{d}) ,
\label{eq:1-phi-1_aux:a}
\end{equation}
with
\begin{equation}
  z_m = \frac{1+y-x}{2}, \qquad d = z_m^2-y ,
\label{eq:1-phi-1_aux:b}
\end{equation}
and
\begin{equation}
  p_0 = \left(\frac{\alpha}{d}\right)^{1/4}\frac{1-x-y}{2}, \qquad
  q_0 = \left(\frac{d}{\alpha}\right)^{1/4} .
\label{eq:1-phi-1_aux:c}
\end{equation}
\end{prop}
\begin{rem}
Errors in the statement of this result as given in \cite{Prellberg} have been fixed, 
in particular the 
sign of the $ q_0 $ term in eq.(\ref{eq:1-phi-1_asymptotic}) and the factor of $ 2 $ in 
the denominator of $ p_0 $ in eq.(\ref{eq:1-phi-1_aux:c}).
\end{rem}

In our case, the parameters are given by
\begin{equation}\label{eq:xy}
\epsilon=\frac{\delta^3}{2},\quad
q={\rm e}^{-\frac{\delta^3}{2}},\quad
y=-{\rm e}^{-\frac{\delta^3}{2}}=-1+\frac{\delta^3}{2},\quad
x=\mp qt=\mp 2i{\rm e}^{-\frac{s}{2}\delta^2+\frac{\delta^3}{2}}=\pm(2i-is\delta^2-i\delta^3+\cdots),
\end{equation}
and unfortunately, they do not match the assumptions of Proposition
\ref{prop:1-phi-1}.  Therefore we have to consider the extension of Proposition
\ref{prop:1-phi-1} for our case, taking care of the multi-valuedness of the
integrand. We fix the branch of $\log$ and fractional power functions as
\begin{displaymath}
  \log z=\ln |z|+i\Arg z,\quad -\pi\leq \Arg z<\pi,
  \qquad
  z^{\frac{m}{n}}=\exp\left(\frac{m}{n}\log z\right)=|z|^\frac{m}{n}{\rm e}^{\frac{m}{n}i\Arg z}.
\end{displaymath}
Note that $\log(XY)=\log X+\log Y$ and $\log(X/Y)=\log X-\log Y$ are valid only
mod $2\pi i$.

Substituting eq.(\ref{eq:xy}) into eq.(\ref{eq:Laplace_qInt}) and expanding the
integrand with respect to $\delta$, we obtain:
\begin{prop}
With the substitutions eq.(\ref{eq:xy}) we have as $\delta\to 0$
\begin{eqnarray}
 {}_1\varphi_1\left({{0}\atop{-q}}~;~q,-qt\right)
&=&
\frac{1}{2\pi i}\int^{\rho+i\infty}_{\rho-i\infty}
{\rm e}^{\frac{2}{\delta^3}g_{-}(z)}
f_{-}(z)~dz\times
\left(\frac{2\pi}{\delta^3}\right)^{\frac{1}{2}}
{\rm e}^{\frac{2}{\delta^3}\left(-\frac{\pi^2}{4}+\frac{\ln 2}{2}\delta^3\right)}\times
\left[1+O(\delta^3)\right], \label{eq:int_-qt}\\
 {}_1\varphi_1\left({{0}\atop{-q}}~;~q,qt\right)
&=&
\frac{1}{2\pi i}\int^{\rho+i\infty}_{\rho-i\infty}
{\rm e}^{\frac{2}{\delta^3}g_{+}(z)}
f_{+}(z)~dz
\times
\left(\frac{2\pi}{\delta^3}\right)^{\frac{1}{2}}
{\rm e}^{\frac{2}{\delta^3}\left(-\frac{\pi^2}{4}+\frac{\ln 2}{2}\delta^3\right)}\times
\left[1+O(\delta^3)\right],\label{eq:int_qt}
\end{eqnarray}
where $0<\rho<1$ and
\begin{eqnarray}
g_{-}(z)&=&{\rm Li}_2(z)-{\rm Li}_2\left(-\frac{1}{z}\right)
+\log\left(2i{\rm e}^{-\frac{s\delta^2}{2}} \right)\log z, \label{eq:g_-qt}\\
f_{-}(z)&=&{\rm e}^{-\frac{1}{2}\log\left(1+\frac{1}{z}\right)-\frac{1}{2}\log(1-z) -\log z},\label{eq:f_-qt}
\end{eqnarray}
\begin{eqnarray}
g_{+}(z)&=&{\rm Li}_2(z)-{\rm Li}_2\left(-\frac{1}{z}\right)
+\log\left(-2i{\rm e}^{-\frac{s\delta^2}{2}} \right)\log z,\label{eq:g_qt}\\
f_{+}(z)&=&{\rm e}^{-\frac{1}{2}\log\left(1+\frac{1}{z}\right)-\frac{1}{2}\log(1-z) -\log z}.\label{eq:f_qt}
\end{eqnarray}
\end{prop}

Let us take the case of $x=-qt$ and apply the saddle point method to
eqs.(\ref{eq:int_-qt}), (\ref{eq:g_-qt}) and (\ref{eq:f_-qt}). The saddle points
$z^{(-)}_1$ and $z^{(-)}_2$ are determined by
\begin{equation}
g_{-}'(z)=-\frac{\log (1-z)}{z}
-\frac{\log\left(1+\frac{1}{z}\right)}{z}
+\frac{\log\left(2i{\rm e}^{-\frac{s\delta^2}{2}} \right)}{z}=0,
\end{equation}
which yields the quadratic equation
\begin{equation}
 z^2+2i{\rm e}^{-\frac{s\delta^2}{2}} z-1=0.
\end{equation}
Therefore the saddle points are given by
\begin{equation}
 z_1^{(-)}=z_m+D,\quad z_2^{(-)}=z_m-D,\quad
z_m=-i{\rm e}^{-\frac{s\delta^2}{2}},
\quad D^{2}=z_m^{2}+1,
\end{equation}
or expanding in terms of $\delta$ we obtain
\begin{equation}
 z_m = -i + \frac{is}{2}\delta^2+\cdots,\quad
D= s^{\frac{1}{2}}\delta - \frac{s^{\frac{3}{2}}}{4}\delta^3+\cdots,
\end{equation}
\begin{equation}\label{eq:z12}
  z_1^{(-)}=-i+s^{\frac{1}{2}}\delta+\frac{is}{2}\delta^2+\cdots,\quad
 z_2^{(-)}=-i-s^{\frac{1}{2}}\delta+\frac{is}{2}\delta^2+\cdots.
\end{equation}
The quantity $\alpha$ is calculated by using eq.(\ref{eq:alpha_beta}) as
\begin{eqnarray*}
\frac{4}{3}\alpha^{\frac{3}{2}} &=& g(z_2^{(-)})-g(z_1^{(-)}) \\
&=&
{\rm Li}_2\left(z_m-D\right)-{\rm Li}_2\left(-\frac{1}{z_m-D}\right)
+\log\left(2i{\rm e}^{-\frac{s\delta^2}{2}}\right)\log(z_m-D)\\
&&-{\rm Li}_2\left(z_m+D\right)+{\rm Li}_2\left(-\frac{1}{z_m+D}\right)
-\log\left(2i{\rm e}^{-\frac{s\delta^2}{2}}\right)\log(z_m+D)\\
&=&2\left[
{\rm Li}_2\left(z_m-D\right)-{\rm Li}_2\left(z_m+D\right)
\right]
+\log\left(2i{\rm e}^{-\frac{s\delta^2}{2}}\right)\left[\log(z_m-D)-\log(z_m+D)\right]\\
&=& \frac{2i}{3}s^{\frac{3}{2}}\delta^3+O(\delta^5)
\end{eqnarray*}
where we have used $\frac{1}{z_m\pm D}=-(z_m\mp D)$. Therefore we conclude
\begin{eqnarray}
% \alpha^{\frac{3}{2}}&=& 2^{-1}{\rm e}^{\frac{\pi i}{2}}s^{\frac{3}{2}}\delta^3+O(\delta^5)\\
 \alpha&=& 2^{-\frac{2}{3}}s{\rm e}^{\frac{\pi  i}{3}}\delta^2+O(\delta^4).
% \alpha^{\frac{1}{2}}&=& 2^{-\frac{1}{3}}s^{\frac{1}{2}}{\rm e}^{\frac{\pi i}{6}}\delta+O(\delta^3)
\end{eqnarray}
We can derive this for $ s $ in the sector
$ -\pi \leq {\rm Arg}(s) < \pi/3 $ but it actually holds without this restriction.
The reason for this is that $ {}_1\varphi_1 $ is an analytic function of 
$ t \in \mathbb{C} $ and therefore of $ s \in \mathbb{C} $. Consequently the leading
term of the expansion of $ {}_1\varphi_1 $ as $ \delta \to 0 $ is analytic with
respect to $ s $ as the remainder terms can be shown to be uniformly bounded in
$ s $ under this limit.
Let us next compute $p_0$ and $q_0$ according to the formula
eq.(\ref{eq:p0_and_q_0}). From the correspondence
\begin{eqnarray*}
z= z_1^{(-)}=-i+s^{\frac{1}{2}}\delta+\frac{is}{2}\delta^2+\cdots
&\longleftrightarrow&
u=u_1^{(-)}=\alpha^{\frac{1}{2}}=2^{-\frac{1}{3}}s^{\frac{1}{2}}{\rm e}^{\frac{\pi i}{6}}\delta+O(\delta^3)\\
z= z_2^{(-)}=-i-s^{\frac{1}{2}}\delta+\frac{is}{2}\delta^2+\cdots
&\longleftrightarrow&
u=u_2^{(-)}=-\alpha^{\frac{1}{2}}=-2^{-\frac{1}{3}}s^{\frac{1}{2}}{\rm e}^{\frac{\pi i}{6}}\delta+O(\delta^3)
\end{eqnarray*}
we obtain $(dz/du)_{z=z_1^{(-)},z_2^{(-)}}$ as
\begin{eqnarray*}
\left(\frac{dz}{du}\right)_{z=z_1^{(-)}}&=&
\frac{\left(\frac{\DS{dz_1}}{\DS{d\delta}}\right)_{\delta\sim 0}}{\left(\frac{\DS{du_1^{(-)}}}{\DS{d\delta}}\right)_{\delta\sim 0}} 
=\frac{s^{\frac{1}{2}}+is\delta+O(\delta^2)}{2^{-\frac{1}{3}}s^{\frac{1}{2}}
{\rm e}^{\frac{\pi i}{6}}+O(\delta^2)}
= 2^{\frac{1}{3}}{\rm e}^{-\frac{\pi i}{6}}\left(1+is^{\frac{1}{2}}\delta+O(\delta^2)\right),\\
\left(\frac{dz}{du}\right)_{z=z_2^{(-)}}&=&
\frac{\left(\frac{\DS{dz_2}}{\DS{d\delta}}\right)_{\delta\sim 0}}{\left(\frac{\DS{du_2^{(-)}}}{\DS{d\delta}}\right)_{\delta\sim 0}} 
=\frac{-s^{\frac{1}{2}}+is\delta+O(\delta^2)}{-2^{-\frac{1}{3}}s^{\frac{1}{2}}
{\rm e}^{\frac{\pi i}{6}}+O(\delta^2)}
= 2^{\frac{1}{3}}{\rm e}^{-\frac{\pi i}{6}}\left(1-is^{\frac{1}{2}}\delta+O(\delta^2)\right).
\end{eqnarray*}
Substituting eq.(\ref{eq:z12}) into eq.(\ref{eq:f_-qt}) we have
\begin{displaymath}
 f(z_1^{(-)})= 2^{-\frac{1}{2}}{\rm e}^{\frac{\pi i}{4}}\left(1-is^{\frac{1}{2}}\delta+\cdots\right),\quad
 f(z_2^{(-)})= 2^{-\frac{1}{2}}{\rm e}^{\frac{\pi i}{4}}\left(1+is^{\frac{1}{2}}\delta+\cdots\right),
\end{displaymath}
from which we obtain
\begin{eqnarray}
p_0&=&\frac{1}{2}\left[
f(z_1^{(-)})\left(\frac{dz}{du}\right)_{z=z_1^{(-)}}+f(z_2^{(-)})\left(\frac{dz}{du}\right)_{z=z_2^{(-)}}\right] 
=2^{-\frac{1}{6}}{\rm e}^{\frac{\pi i}{12}}+O(\delta^2), \\
q_0&=&\frac{1}{2\alpha^{\frac{1}{2}}}
\left[
f(z_1^{(-)})\left(\frac{dz}{du}\right)_{z=z_1^{(-)}}-f(z_2^{(-)})\left(\frac{dz}{du}\right)_{z=z_2^{(-)}}
\right] =0 +O(\delta).
\end{eqnarray}
We compute $\beta$ by using eq.(\ref{eq:alpha_beta}) as
\begin{eqnarray}
 2\beta&=&g(z_2^{(-)})+g(z_1^{(-)})\nonumber\\
&=&{\rm Li}_2\left(z_m-D\right)-{\rm Li}_2\left(-\frac{1}{z_m-D}\right)
+\log\left(2i{\rm e}^{-\frac{s\delta^2}{2}}\right)\log(z_m-D)\nonumber\\
&+&{\rm Li}_2\left(z_m+D\right)-{\rm Li}_2\left(-\frac{1}{z_m+D}\right)
+\log\left(2i{\rm e}^{-\frac{s\delta^2}{2}}\right)\log(z_m+D)\nonumber\\
&=&\log\left(2i{\rm e}^{-\frac{s\delta^2}{2}}\right)\left[\log(z_m-D)+\log(z_m+D)\right],\label{eq:beta}
\end{eqnarray}
which yields
\begin{equation}
  \beta=-\frac{\pi i}{2} \ln 2 + \frac{\pi^2}{4}+\frac{\pi i}{4}s\delta^2+O(\delta^4).
\end{equation}
\begin{rem}
The multi-valuedness of the integrand has a critical effect in the calculation of
$\beta$. One might compute $\beta$ from eq.(\ref{eq:beta}) as
\begin{eqnarray*}
2\beta&=&\log\left(2i{\rm  e}^{-\frac{s\delta^2}{2}}\right)\left[\log(z_m-D)+\log(z_m+D)\right]
=\log\left(2i{\rm e}^{-\frac{s\delta^2}{2}}\right)\log(z_m^2-D^2)\\
&=&\log\left(2i{\rm e}^{-\frac{s\delta^2}{2}}\right)\log(-1)=-\pi i(\ln 2 + \frac{\pi i}{2}-\frac{s\delta^2}{2}),
\end{eqnarray*}
but the second equality does not hold in general (in this case it is
accidentally correct). In fact, the same procedure for the case of $x=qt$
yields wrong result.
\end{rem}

Let us finally consider the image of integration path $C: z=\rho+it$
($-\infty<t<\infty$) in the $u$-plane. 
From the identity of dilogarithm\cite{Kirillov,Erdelyj}
\begin{equation}
 {\rm Li}_2(z)=-{\rm Li}_2(\frac{1}{z})-\frac{1}{2}(\log z)^2+\pi i\log z+\frac{\pi^2}{3},
\end{equation}
we see for $t\to\pm \infty$
\begin{eqnarray}
 g_{-}(\rho+it)&\sim&{\rm Li}_2(\rho+it)=-{\rm Li}_2(\frac{1}{\rho+it})-\frac{1}{2}(\log (\rho+it))^2+\pi i\log
 (\rho+it)+\frac{\pi^2}{3}\nonumber\\
&\sim& -\frac{1}{2}(\log |t|)^2 \pm \frac{\pi i}{2}\log|t|,\quad t\to \pm \infty,
\end{eqnarray}
therefore
\begin{displaymath}
 u(\rho+it)\sim (3g_{-}(\rho+it))^{\frac{1}{3}}
\sim {\rm e}^{\pm \frac{\pi i}{3}}\left(\frac{3}{2}(\log|t|)^2\right)^{\frac{1}{3}},
\quad t\to \pm \infty.
\end{displaymath}
This gives the integration path $C'$ as 
$\infty{\rm e}^{-\frac{\pi i}{3}}\to O\to \infty{\rm e}^{\frac{\pi i}{3}}$, which implies 
$V(\lambda)={\rm Ai}(\lambda)$. Closer investigation shows that the mapping given by
eq.(\ref{eq:g_and_u}) is regular and 1:1 in the domain including $C'$.

Collecting the above results and performing similar calculations for the case of
$x=qt$, we obtain the following asymptotic expansions from eq.(\ref{eq:expansion_Airy}):
\begin{prop}
With the substitutions eq.(\ref{eq:cont_params}) we have as $\delta\to 0$
\begin{eqnarray}
 {}_1\varphi_1\left({{0}\atop{-q}}~;~q,-qt\right)
&=&2\pi^{\frac{1}{2}}\delta^{-\frac{1}{2}}
{\rm e}^{-\frac{\pi i}{\delta^3} \ln 2 +\frac{\pi i}{2\delta}s+\frac{\pi i}{12}}
\left[{\rm Ai}(s{\rm e}^{\frac{\pi i}{3}})+O(\delta^2)\right], \label{eq:Airy-limit1}\\
 {}_1\varphi_1\left({{0}\atop{-q}}~;~q,qt\right)
&=& 
2\pi^{\frac{1}{2}}\delta^{-\frac{1}{2}}
{\rm e}^{\frac{\pi i}{\delta^3} \ln 2 -\frac{\pi i}{2\delta}s-\frac{\pi i}{12}}
\left[{\rm Ai}(s{\rm e}^{-\frac{\pi i}{3}})+O(\delta^2)\right],\label{eq:Airy-limit2}
\end{eqnarray}
for $s$ in any compact domain of $\mathbb{C}$.
\end{prop}

Now we are in a position to deduce the limit of the general solution to 
eq.(\ref{eq:qAiry}). From eqs.(\ref{vDefn}) and (\ref{eq:gen_sol_of_qAiry}) we note that
\begin{equation}
v(t) = A~{\rm e}^{-\frac{\pi i}{2}\frac{\log t}{\log q}}
       {}_1\varphi_1\left(\begin{array}{c}0 \\-q \end{array};q,-qt\right)
       +B~{\rm e}^{\frac{\pi i}{2}\frac{\log t}{\log q}}
       {}_1\varphi_1\left(\begin{array}{c}0 \\-q \end{array};q,qt\right) ,
\end{equation}
for arbitrary $q$-periodic functions $ A,B $. Observing that
\begin{equation}
{\rm e}^{\pm \frac{\pi i}{2} \frac{\log t}{\log q}}=
   {\rm e}^{\frac{1}{\delta^3}\left(\mp\pi i\ln 2\mp\frac{\pi^2}{2}\right)\pm\frac{\pi i}{2\delta}s}
   \times[1+O(\delta)],
\end{equation}
we find that the $ s $-dependence in the exponential 
pre-factors of eqs.(\ref{eq:Airy-limit1})  and (\ref{eq:Airy-limit2}) cancels
exactly. Therefore we finally arrive at the desired result:
\begin{thm}
We have as $ \delta \to 0 $
\begin{equation}
 v(t) = 2\sqrt{\frac{\pi}{\delta}}
    \left[ A~{\rm e}^{\frac{\pi^2}{2\delta^3}+\frac{\pi i}{12}}{\rm Ai}({\rm e}^{\frac{\pi i}{3}}s)
          +B~{\rm e}^{-\frac{\pi^2}{2\delta^3}-\frac{\pi i}{12}}{\rm Ai}({\rm e}^{-\frac{\pi i}{3}}s)
    \right]\times [1+O(\delta)],
\end{equation}
for $ s $ in any compact domain of $ \mathbb{C} $ and constants $ A,B \in \mathbb{C} $.
\end{thm}

\section{Concluding Remarks}
In this article we have considered the $q$-Painlev\'e equation dP($A_6^{(1)}$),
eq.(\ref{eq:qp2}), and constructed its classical solutions having a determinantal form
with basic hypergeometric function elements. We have also discussed the continuous 
limit to P$_{\rm II}$. In particular, we have shown that hypergeometric functions 
${}_1\varphi_1\left(\begin{array}{c}0 \\-q \end{array};q,\mp qt\right)$ actually
reduce to the Airy functions ${\rm Ai}({\rm e}^{\pm \frac{\pi i}{3}})$ by applying
a generalization of the saddle point method to their integral representations. 

We first remark that P$_{\rm II}$ eq.(\ref{eq:p2}) admits rational
solutions for $\eta=2N$ ($N\in\mathbb{Z}$) which can be expressed in
terms of a specialization of 2-core Schur functions\cite{KO:p2}. Such
solutions are obtained by applying B\"acklund transformations to the
simple rational solution that is fixed by the Dynkin diagram
automorphism. One would expect similar rational solutions for
eq.(\ref{eq:qp2}), however Masuda has shown that there is no rational
solution fixed by the corresponding Dynkin diagram automorphism
\cite{Masuda:private}. This implies that it is not appropriate to regard
eq.(\ref{eq:qp2}) simply as a ``$q$-analog of P$_{\rm II}$''.  

Secondly, we observe an asymmetry in the structure of determinant
formula eq.(\ref{tau:qp2}); the shifts in $t$ of the entries are different between
the horizontal and the vertical directions. It might be natural to regard this
structure as originating from the asymmetry of the root lattice. Actually
in other cases where this situation arises, such as the
$A_5^{(1)}$ surface(``$q$-P$_{\rm IV}$'' and ``$q$-P$_{\rm III}$'',
$(A_2+A_1)^{(1)}$-symmetry)\cite{KNY:qp4,KK:qp3},
$A_4^{(1)}$ surface(``$q$-P$_{\rm V}$'', $A_4^{(1)}$-symmetry)\cite{HK:qp5} or 
$A_3^{(1)}$ surface(``$q$-P$_{\rm VI}$'', $D_5^{(1)}$-symmetry)\cite{Sakai:qp6sol}, the
determinant structure is symmetric.  Such an asymmetric structure of the
determinant is known for several discrete Painlev\'e equations - one example is
the ``standard'' discrete Painlev\'e II
equation\cite{RGH:dP,KOSGR:dP2}
\begin{equation}\label{eq:standard_dP2}
 x_{n+1}+x_{n-1}=\frac{(an+b)x_n+c}{1-x_n^2},
\end{equation}
where $a,b,c$ are parameters. Equation (\ref{eq:standard_dP2}) may be
regarded as a special case of the ``asymmetric'' discrete Painlev\'e II
equation\footnote{The word ``asymmetric'' comes from the terminology of
the Quispel-Roberts-Thompson mapping which is known as the fundamental
family of second-order integrable mappings. It has nothing to do
with the asymmetric structure of the determinant formula of their
solutions.}\cite{ROSG,Ohta:RIMS,RGTT}
\begin{equation}\label{eq:asymmetric_dP2}
 x_{n+1}+x_{n-1}=\frac{(an+b)x_n+c+d(-1)^n}{1-x_n^2},
\end{equation}
when $d=0$,
which is actually a discrete Painlev\'e equation associated with the
$D_5^{(1)}$ surface and arises as a B\"acklund transformation of the Painlev\'e
V equation (P$_{\rm V}$). Therefore the hypergeometric solutions for
(\ref{eq:asymmetric_dP2}) are expressible in terms of the Whittaker
function and are the same as those for P$_{\rm V}$\cite{RGTT}. However, the
hypergeometric solutions for eq.(\ref{eq:standard_dP2}) are quite
different; the relevant hypergeometric function is the parabolic
cylinder function and the determinant structure has the same asymmetry
as that for dP($A_6^{(1)}$) eq.(\ref{eq:qp2}). A similar structure is 
known for the ``standard'' discrete ($q$-)Painlev\'e III equation
(dP$_{\rm III}$)\cite{RGH:dP,KOS:dP3}
\begin{equation}\label{eq:standard_dP3}
 \frac{x_{n+1}x_{n-1}}{d_1d_2}=\frac{(x_n-c_1q^n)(x_n-c_2q^n)}{(x_n-d_1)(x_n-d_2)},
\end{equation}
and the $q$-Painlev\'e VI equation ($q$-P$_{\rm VI}$) \cite{Jimbo-Sakai:qp6,ROSG},
\begin{equation}\label{eq:qP6}
\begin{array}{l}
\smallskip
 y_ny_{n+1}=\dfrac{a_3a_4(z_{n+1}-b_1q^n)(z_{n+1}-b_2q^n)}
{(z_{n+1}-b_3)(z_{n+1}-b_4)},\\
\smallskip
 z_nz_{n+1}=\dfrac{b_3b_4(y_{n}-a_1q^n)(y_{n}-a_2q^n)}
{(y_{n}-a_3)(y_{n}-a_4)},
\end{array}\quad
\dfrac{b_1b_2}{b_3b_4}=q\dfrac{a_1a_2}{a_3a_4},
\end{equation}
where $a_i$, $b_i$ ($i=1,2,3,4$), $c_j$ and $d_j$ ($j=1,2$) are
parameters. The hypergeometric solutions for eq.(\ref{eq:qP6}) are given
by the basic hypergeometric series ${}_2\varphi_1$ \cite{Sakai:qp6sol} while those for
eq.(\ref{eq:standard_dP3}) are given by Jackson's $q$-Bessel function\cite{KOS:dP3}.

Thus the results of our study sheds some light on the ``degenerated''
equations such as eqs.(\ref{eq:standard_dP2}) or (\ref{eq:standard_dP3}). 
They are not just special cases of the original ``generic''
equations. Our results imply that putting $d=0$ in
eq.(\ref{eq:asymmetric_dP2}) is not just killing the ``parity'', but
causes qualitative change of the root lattice which in turn results in different
hypergeometric solutions and an asymmetry of the determinant formula. Therefore
they should be studied independently of the ``generic'' discrete Painlev\'e
equations in the Sakai's classification\cite{Sakai:elliptic}.\\

\noindent\textbf{Acknowledgement} We acknowledge Profs. Takashi Aoki,
Takahiro Kawai, Tatsuya Koike and Yoshitsugu Takei for their interests
in this work and valuable discussions. We also thank Profs. Tetsu
Masuda, Masatoshi Noumi, Yasuhiro Ohta and Yasuhiko Yamada for their
continuous encouragement and discussions. NSW acknowledges support from
the Australian Research Council.

\end{document}